\documentclass[aps,pre,twocolumn,superscriptaddress,groupedaddress]{revtex4}

\usepackage{graphicx}  
\usepackage{dcolumn}   
\usepackage{bm}        
\usepackage{amssymb}   
\usepackage{amsmath, amsfonts}
\usepackage{xcolor}

\begin{document}



\title{Chaotic wave packet spreading in two-dimensional disordered nonlinear lattices}
\author{B.~Many Manda}
\affiliation{Department of Mathematics and Applied Mathematics, University of Cape Town, Rondebosch, 7701, Cape Town, South Africa}
\author{B.~Senyange}
\affiliation{Department of Mathematics and Applied Mathematics, University of Cape Town, Rondebosch, 7701, Cape Town, South Africa}
\author{Ch.~Skokos}
\email{haris.skokos@uct.ac.za}
\thanks{Corresponding author.}
\affiliation{Department of Mathematics and Applied Mathematics, University of Cape Town, Rondebosch, 7701, Cape Town, South Africa}

\date{\today}

\begin{abstract}
We reveal the generic characteristics of wave packet delocalization in two-dimensional nonlinear disordered lattices by performing extensive numerical simulations in two basic disordered models: the Klein-Gordon system and the discrete nonlinear Schr\"{o}dinger equation. We find that in both models (a) the wave packet's second moment asymptotically evolves as $t^{a_m}$ with $a_m \approx 1/5$ ($1/3$) for the weak (strong) chaos dynamical regime, in agreement with previous theoretical predictions [S.~Flach, Chem.~Phys.~{\bf 375}, 548 (2010)], (b) chaos persists, but its strength decreases in time $t$ since the finite time maximum Lyapunov exponent $\Lambda$ decays as  $\Lambda \propto t^{\alpha_{\Lambda}}$, with $\alpha_{\Lambda} \approx -0.37$ ($-0.46$) for the weak (strong) chaos case, and (c) the deviation vector distributions show the wandering of localized chaotic seeds in the lattice's excited part, which induces  the wave packet's thermalization. We also propose a dimension-independent scaling between the wave packet's spreading and chaoticity, which allows the prediction of the obtained $\alpha_{\Lambda}$ values.
\end{abstract}

\maketitle

\section{\label{par:introduction} Introduction}

The normal modes (NMs) of disordered linear lattices are spatially localized for strong enough disorder, and consequently any initial compact wave packet  stays localized forever.
This pioneering theoretical result was obtained by
Anderson and is  referred to as {\it Anderson localization} (AL)~\cite{AL}.
Several experimental manifestations of  AL have  been reported to date~\cite{SBFS07RDFFFZMMI08BJZBHLCSBA08}. However, what happens to  AL in the presence of nonlinearity  is still an open question, which has been discussed widely~\cite{dis_many,KKFA08,GS09,FKS09,SKKF09,MAPS09,OPH09,KF10, LBKSF10,F10,BLSKF11,B11,B12,LBF12,SGF13,M14,SMS18,VFF19}. Two models have been at the center of these studies: the disordered Klein-Gordon (DKG) lattice of coupled anharmonic oscillators and the disordered discrete nonlinear Schr\"odinger equation (DDNLS).
For both models, it was found that nonlinearity eventually destroys AL, leading  to a slow subdiffusive spreading of  wave packets, whose second moment grows in time $t$ as $t^{a_m}$ ($0<a_m<1$)~\cite{dis_many,GS09,FKS09,SKKF09,LBKSF10,F10,BLSKF11,LBF12}.
In particular, an {\it asymptotic} spreading regime called~\textit{`weak chaos'} where $a_m = 1/(1 + 2d)$ ($d$ being the lattice spatial dimension) was identified~\cite{FKS09,SKKF09,F10}, while an {\it intermediate} spreading regime named~\textit{`strong chaos'}, with $a_m = 1/(1 + d)$, may also occur~\cite{LBKSF10,F10,BLSKF11}.

The wave packet spreading in nonlinear disordered lattices is a chaotic process induced by the systems' non-integrability and resonances between NMs~\cite{SKKF09,KF10,F10}. Such  deterministic chaotic processes result to the randomization and thermalization of wave packets~\cite{MAPS09,B11,B12,SGF13,M14,SMS18}.
The computation of the finite time maximum Lyapunov exponent (MLE) $\Lambda$ for initially localized excitations in  one-dimensional (1D) lattices~\cite{SGF13,SMS18} showed that the  wave packet's chaoticity is characterized by a positive but decaying  MLE. Furthermore, the evolution of the deviation vector associated to $\Lambda$ indicated the existence of chaotic seeds which randomly wander inside the wave packet ensuring the chaotization of the excited degrees of freedom.

Although the dynamics of 1D  lattices has been studied extensively, less numerical work has been done for 2D systems. One of the main obstacles there is the very large computational effort required for the long time simulation of these models (especially of the 2D DDNLS system). In~\cite{GS09} the wave packet spreading in the 2D DDNLS model for the weak chaos regime was studied up to $t=10^6$ time units, while in~\cite{SDNLD18} a similar model, including also non-diagonal nonlinear terms, was considered. In both cases statistical analyses over a few disorder realizations were performed~\cite{comDDNLS}. In~\cite{LBF12}  results for  times up to $t=10^8$ with statistics over $400$ realizations were reported, but only for the 2D DKG model. There nonlinear terms with different exponents were also considered. For the typical DKG system with quartic nonlinearities only the weak chaos regime was investigated, probably because the strong chaos case, which is characterized by faster spreadings, would require the, computationally demanding, integration of larger lattices.

Here we focus our attention on the 2D DKG and DDNLS models with quartic nonlinearities. We not only study the characteristics of wave packet spreading for both the weak and strong chaos regimes, but also analyze in depth their chaotic behavior through the computation of their MLE and the associated deviation vector distributions (DVDs), as was done in~\cite{SGF13,SMS18} for their 1D counterparts.

The paper is organized as follows. In Sec.~\ref{par:model_and_Computational} we present the two Hamiltonian models we consider in our study, along with the various quantities we use in order to analyze their dynamical behavior. In addition, we provide information about our numerical computations. In Sec.~\ref{par:results} we present our numerical findings about the chaotic behavior of the DKG and the DDNLS models, while in Sec.~\ref{par:discussion_conclusion} we summarize our findings and discuss our conclusions.

\section{\label{par:model_and_Computational} Models and Computational Aspects}

The Hamiltonian of the 2D DKG system~\cite{LBF12, SS18} in canonical coordinates $q_{l, m}$ (positions) and $p_{l, m}$ (momenta) is
\begin{eqnarray}
\label{eq:hamiltonian_dkg}
H_K = \sum _{l, m}
\bigg\{ \frac{p_{l, m}^2}{2}  & + &  \frac{\epsilon_{l, m} }{2}q_{l, m}^2 + \frac{q_{l, m}^4}{4} + \frac{1}{2W}
 \\
& \times &  \left[\left(q_{l, m+1} - q_{l, m}\right)^2  + \left(q_{l+1, m} - q_{l, m}\right)^2  \right] \bigg\},
\nonumber
\end{eqnarray}
with $\epsilon_{l, m}$ being uncorrelated parameters uniformly distributed on the interval $[1/2, 3/2]$. The Hamiltonian of the 2D DDNLS model~\cite{GS09,LBF12,DMTS19} in real canonical coordinates $q_{l, m}$ and $p_{l, m}$ reads
\begin{eqnarray}
\label{eq:hamiltonian_ddnls_real}
&H_D & = \sum_{l, m} \bigg[ \frac{\hat{\epsilon} _{l, m}}{2}  \left( q_{l, m}^2 + p_{l, m}^2 \right)  +  \frac{\beta}{8} \left( q_{l, m}^2 + p_{l, m}^2 \right)^2 - \\
&	& \left( q_{l, m + 1} q_{l, m} + q_{l + 1, m} q_{l, m} + p_{l, m + 1} p_{l, m} + p_{l + 1, m} p_{l, m} \right) \bigg], \nonumber
\end{eqnarray}
where $\hat{\epsilon}_{l, m}$ are random numbers uniformly drawn in the interval $\left[-W/2, W/2\right]$ and $\beta \ge 0$ is the nonlinear coefficient. In Eqs.~\eqref{eq:hamiltonian_dkg} and~\eqref{eq:hamiltonian_ddnls_real} $l$ and $m$ are integer indices, $W$ represents the  disorder strength and fixed boundary conditions are imposed.
The system's evolution  conserves the Hamiltonian value (also referred as energy) $H=H_K$~\eqref{eq:hamiltonian_dkg} [$H_D$~\eqref{eq:hamiltonian_ddnls_real}] for the DKG [DDNLS] model. The DDNLS system has an additional integral of motion: the {\it norm} $S = \sum_{l, m}(q_{l, m}^2 + p_{l, m}^2)/2$. We define for the DKG model the normalized energy density distribution $\xi _{l, m} = h_{l, m}/H_{K}$~\cite{LBF12, SS18}, where
\begin{eqnarray}
\label{eq:hlm}
&h_{l, m}=& \frac{p_{l, m}^2}{2} + \frac{\epsilon_{l, m}}{2}q_{l, m}^2 + \frac{q_{l, m}^4}{4} +
 \frac{1}{4W}
  \\
   & & \times \left[\left(q_{l, m} - q_{l-1, m}\right)^2 + \left(q_{l, m} - q_{l, m-1}\right)^2 + \right. \nonumber
    \\
     & &\left.\left(q_{l, m+1} - q_{l, m}\right)^2  + \left(q_{l+1, m} - q_{l, m}\right)^2\right], \nonumber
\end{eqnarray}
is the energy of site $(l,m)$, while for the DDNLS system the normalized norm density distribution $\xi _{l, m} =s_{l, m}/S$~\cite{LBF12,DMTS19}, with
\begin{equation}
\label{eq:slm}
s_{l, m} = \frac{q_{l, m}^2 + p_{l, m}^2}{2}.
\end{equation}

We follow the evolution of a compact square excitation of side $L$  in the middle of the lattice, so that all initially excited sites of the DKG (DDNLS) system have the same $h_{l,m}$ ($s_{l,m}$) value. We also investigate the systems' chaoticity by computing the finite time MLE~\cite{BGGS80b,S10}
\begin{equation}
\label{eq:mle_new}
\Lambda (t) = \frac{1}{t} \ln \left[ \frac{\| \boldsymbol{w} (t)\|}{ \| \boldsymbol{w} (0) \|} \right],
\end{equation}
where $\boldsymbol{w} (0)$ and $\boldsymbol{w} (t)$ is respectively a deviation vector to the systems' considered orbit at times $t=0$ and $t > 0$.
Here $\| \cdot \|$ represents the usual Euclidian norm.
For regular orbits $\Lambda$ tends to zero as $\Lambda \propto t^{-1}$~\cite{BGGS80b,S10}, otherwise the orbit is considered chaotic.
The deviation vector's  coordinates are small perturbations $\delta q_{l, m} (t)$, $\delta p_{l, m}(t)$, whose  evolution is governed by the so-called {\it variational equations} (see e.g.~\cite{S10}).
We also compute the DVD~\cite{SGF13,SMS18}
\begin{equation}
\label{eq:DVD_new}
\xi^D _{l, m} = \frac{\delta q_{l, m} ^2 + \delta p_{l, m} ^2}{\sum_{l, m} (\delta q_{l, m} ^2 + \delta p_{l, m} ^2)}.
\end{equation}
For all  mentioned distributions, we calculate the second moment
\begin{equation}
\label{eq:m2_new}
m_2^{(D)}=\sum_{l, m} \left\| \boldsymbol{r}_{l, m}^{(D)} - \overline{\boldsymbol{r}}_{l, m}^{(D)}\right\|  ^2 \xi_{l, m}^{(D)},
\end{equation}
which quantifies the distribution's extent, and the participation number
\begin{equation}
\label{eq:P_new}
P^{(D)} = \frac{1}{\sum_{l, m} \left( \xi_{l, m}^{(D)} \right)^2},
\end{equation}
which measures the number of highly excited sites, where $\boldsymbol{r}_{l, m}^{(D)} = (l^{(D)}, m^{(D)})^T$ and $\overline{\boldsymbol{r}}_{l, m}^{(D)} = (\overline{l}^{(D)}, \overline{m}^{(D)})^T = (\sum_{l, m} l \xi_{l, m}^{(D)},\sum_{l, m} m \xi_{l, m}^{(D)})^T$ is the distribution's center, with $(^T)$ denoting the matrix transpose and $(^{(D)})$ referring to the DVD.

We implement the $ABA864$ symplectic integrator~\cite{BCFLMM13,SS18} for the evolution of the  DKG model along with the~\textit{tangent map method} for the integration of its variational equations~\cite{SG10GS11GES12}, and the $s11ABC6$ scheme~\cite{SGBPE14GMS16,DMTS19} for the DDNLS system.
Typically, we perform  simulations up to a final time $t_f \approx 10^6-10^8$.
In order to exclude finite size effects lattice sizes up to $450\times 450$ were considered, which were always much larger than the NMs' average participation number. This quantity decreases when $W$ grows as is seen in the inset of Fig.~1 of~\cite{LBF12}, where it was called `localization volume'~\cite{comment_NM}.
The used integration time steps $\tau \approx 0.1- 1.15$ result to a good conservation of the systems' integrals of motion as the relative energy (norm) error was always kept below $10^{-3}$ ($10^{-2}$). We average the values of an observable $Q$ over  $50$ disorder realizations (denoting by $\langle Q \rangle$ the obtained average value) and evaluate the related local variation $\alpha_Q = d \langle \log_{10}  Q \rangle / d \log_{10} t$ through a regression method~\cite{CD88} as in~\cite{LBKSF10,BLSKF11,LBF12,SGF13,SMS18}.

\section{\label{par:results} Results}

\begin{figure}
\includegraphics[width=\columnwidth]{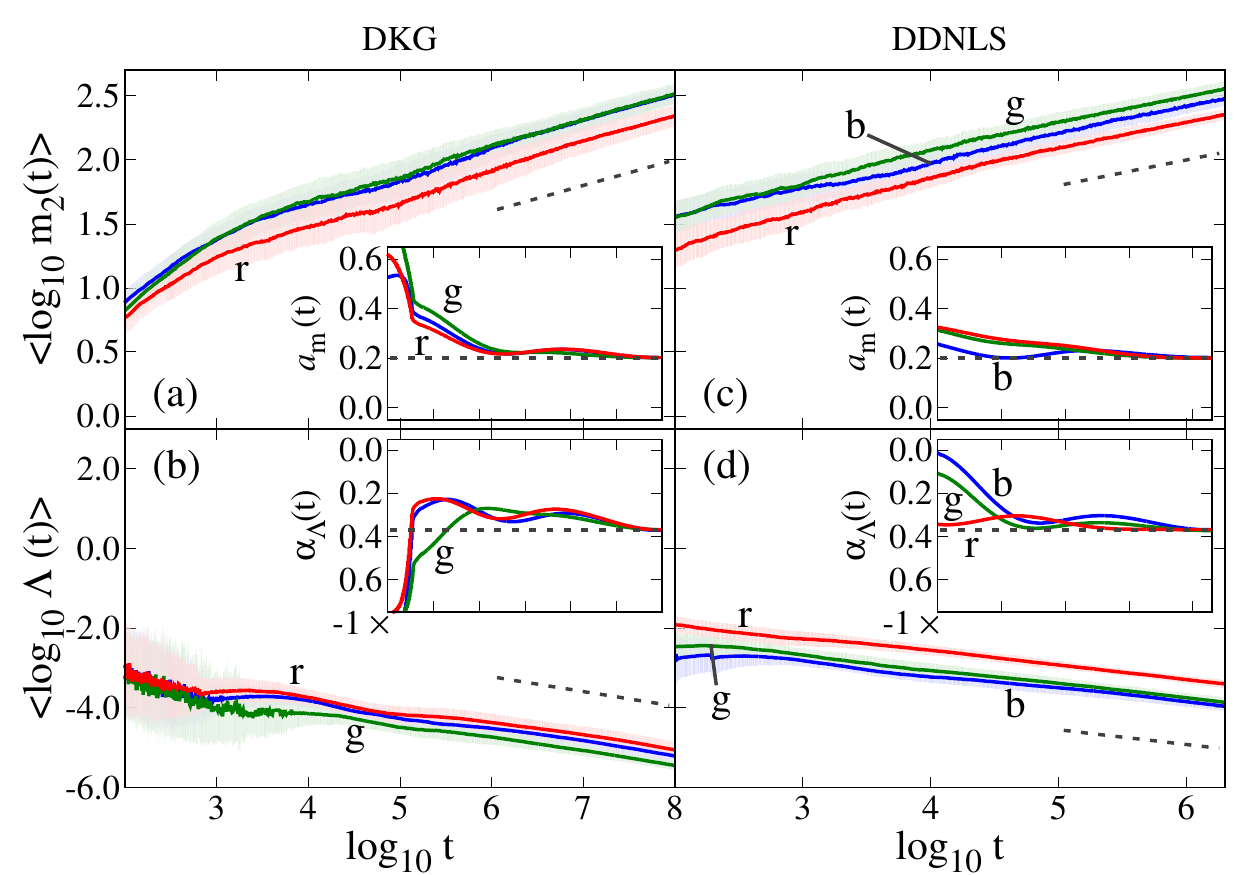}
\caption{\label{fig:sec_mom_mle_weak} Weak chaos. Averaged results of the  evolution of $m_2(t)$ [(a) and (c)] and  $\Lambda (t)$ [(b) and (d)] for (a) and (b) the DKG, and (c) and (d) the DDNLS model.
The presented cases are $W1_K$, $W2_K$ and $W3_K$ for the DKG, and $W1_D$, $W2_D$ and $W3_D$ the DDNLS model [blue (b), green (g) and red (r), respectively for both models]. Shaded areas indicate one standard deviation. Insets: the associated derivatives $a_{m}(t)$ [(a) and (c)] and $\alpha_{\Lambda}(t)$ [(b) and (d)]. The straight dashed lines indicate $a_{m} = 0.2$ [(a) and (c)] and $\alpha_{\Lambda} = -0.37$ [(b) and (d)].}
\end{figure}

Initially, we study the weak chaos regime. For the DKG system we consider the  cases
$W=10$, $L=3$, $h_{l, m} = 0.0085$ (Case $W1_K$), $W=10$, $L=1$, $h_{l, m} = 0.05$ (Case $W2_K$) and $W=11$, $L=2$, $h_{l, m} = 0.0175$ (Case $W3_K$). For the DDNLS system we set
$W=10$, $L=2$, $\beta=0.15$, $s_{l, m}=1$, $H_D \in [-1.9, 0.73]$ (Case $W1_D$), $W=10$, $L=1$, $\beta=0.92$, $s_{l, m}=1$, $H_D =0.5$ (Case $W2_D$) and $W=12$, $L=1$, $\beta=1.75$, $s_{l, m}=1$, $H_D =0.5$ (Case $W3_D$).  We note that for single site excitations ($L=1$) we keep the value $\hat{\epsilon}_{l, m}$ of the initially excited site constant in all disorder realizations so that all cases have the same $H_D$, while for $L>1$ $H_D$ depends on the particular realization. Since the DDNLS system admits two integrals of motion, and always $s_{l, m}$ is fixed, we take particular care so that the used $H_D$ values correspond to the Gibbsian region of the energy-norm density space~\cite{F16TYDF18}. The evolution of $m_2(t)$ both for the DKG [Fig.~\ref{fig:sec_mom_mle_weak}(a)] and the DDNLS model [Fig.~\ref{fig:sec_mom_mle_weak}(c)] clearly shows an asymptotic power law increase $m_2\propto t^{a_m}$ with  $a_m \approx 0.2$~\cite{com1}, in agreement to the theoretically obtained value $a_m=1/5$~\cite{F10}. This value was also retrieved in~\cite{LBF12}, but only for one DKG case. The chaotic nature of all these weak chaos cases becomes evident from Figs.~\ref{fig:sec_mom_mle_weak}(b) and (d) where the evolution of $\Lambda (t)$ is shown. For both models we find an asymptotic decrease $\Lambda \propto t^{\alpha_{\Lambda}}$, with $\alpha_{\Lambda} \neq -1$, similarly to what has been observed for  1D lattices~\cite{SGF13,SMS18}. In particular, $\alpha_{\Lambda}$ converges around $\alpha_{\Lambda}=-0.37$ for all cases.

\begin{figure}
\includegraphics[width=\columnwidth]{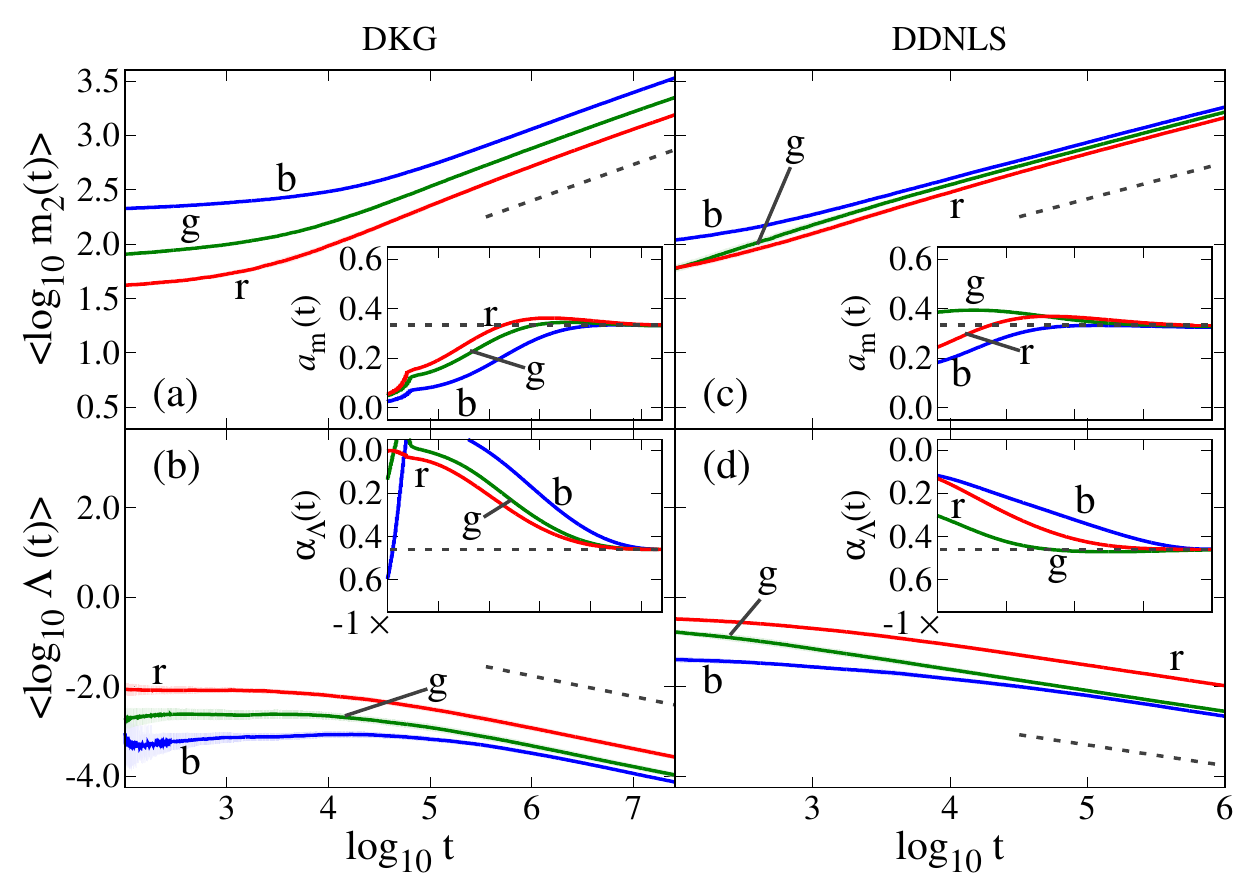}
\caption{\label{fig:sec_mon_mle_strong}
Strong chaos. Similar to Fig.~\ref{fig:sec_mom_mle_weak}. The presented cases are $S1_K$, $S2_K$ and $S3_K$ for the DKG, and $S1_D$, $S2_D$ and $S3_D$  the DDNLS model [blue (b), green (g) and red (r), respectively for both models]. The straight dashed lines indicate $a_{m} = 0.33$ [(a) and (c)] and $\alpha_{\Lambda} = -0.46$ [(b) and (d)].}
\end{figure}

We also investigate the strong chaos regime, which was not studied before for systems \eqref{eq:hamiltonian_dkg} and \eqref{eq:hamiltonian_ddnls_real}.
For the DKG system we consider the  cases
$W=9$, $L=35$, $h_{l, m} = 0.006$ (Case $S1_K$), $W=10$, $L=21$, $h_{l, m} = 0.0135$ (Case $S2_K$) and $W=12.5$, $L=15$, $h_{l, m} = 0.035$ (Case $W3_K$), while for the DDNLS system we set
$W=10.5$, $L=21$, $\beta=0.145$, $s_{l, m}=1$, $H_D \in [0, 61.74]$ (Case $S1_D$), $W=11$, $L=10$, $\beta=0.68$, $s_{l, m}=1$, $H_D \in [-6, 3.5]$  (Case $S2_D$) and $W=14$, $L=15$, $\beta=6$, $s_{l, m}=0.12$, $H_D \in [0, 0.75]$ (Case $S3_D$). The evolution of $m_2 (t)$ for these cases [Figs.~\ref{fig:sec_mon_mle_strong}(a) and (c)] shows again that eventually $m_2\propto t^{a_{m}}$, but with $a_m \approx 0.33$~\cite{com2}. These results confirm the validity of the theoretical analysis of~\cite{F10} where the value $a_m = 1/3$ was predicted.
The computation of $\Lambda (t)$ for all these DKG [Fig.~\ref{fig:sec_mon_mle_strong}(b)] and DDNLS [Fig.~\ref{fig:sec_mon_mle_strong}(d)] strong chaos cases show again a power law decay $\Lambda \propto t^{\alpha_{\Lambda}}$, but now $\alpha_{\Lambda} \approx -0.46$.

Similarly to 1D systems~\cite{SGF13,SMS18}, the results of Figs.~\ref{fig:sec_mom_mle_weak} and~\ref{fig:sec_mon_mle_strong} show, for both the weak and strong chaos regimes, subdiffusive spreading which remains chaotic up to the largest simulation times. The systems become less chaotic as $\Lambda$ decreases in time, but since this decrease remains always different from the $\Lambda \propto t^{-1}$ law observed for regular motion [$\Lambda \propto t^{-0.37} \gg t^{-1}$ and $\Lambda \propto t^{-0.46} \gg t^{-1}$, for the weak and strong chaos regimes respectively] we do not find any signs of a crossover to regular dynamics as it was speculated in~\cite{JKA10A11}. The time evolution of $\Lambda$ can be understood in a similar way as in 1D systems~\cite{SGF13,SMS18}. As time grows the constant total energy (norm) of the DKG (DDNLS) system is shared among more degrees of freedom as additional lattice sites are excited. In this way the energy (norm) density of the excited sites, which quantifies the effective strength of nonlinearity, diminishes leading to a decrease of chaos strength, which is reflected in the power law decay of $\Lambda$.

Following~\cite{SGF13,SMS18} we find that the wave packets' chaotization is done fast enough to support its spreading since the Lyapunov time $T_L = 1 / \Lambda$, which determines a time scale for the systems' chaotization, remains always smaller than the characteristic spreading time $T_D = 1 / D$ (with $D$ being the momentary diffusion coefficient defined through $m_2 \sim Dt$). In particular, the ratio of these time scales
\begin{equation} \label{eq:scales}
\frac{T_D}{T_L } \sim t^{1 + \alpha_{\Lambda} - a_m},
\end{equation}
becomes $T_D / T_L \sim t^{0.43}$ ($t^{0.21}$) for the weak (strong) chaos regime. The fact that these ratios are very close to the ones observed for the 1D counterparts of systems \eqref{eq:hamiltonian_dkg} and \eqref{eq:hamiltonian_ddnls_real}~\cite{SGF13,SMS18}, i.e.~$T_D / T_L \sim t^{0.42}$ ($t^{0.2}$) for the weak (strong) chaos regime, strongly suggests that {\it nonlinear interactions of the same nature are responsible for the chaotic wave packet spreading in one and two spatial dimensions}.

Investigating further the relation between 1D and 2D systems we note that, for both dynamical regimes, the rate of spreading in 2D models (quantified by the exponent $a_m$) is smaller than in  the 1D case. Moreover, 2D systems are less chaotic than their 1D counterparts as their $\Lambda$ decreases faster (i.e.~smaller, negative $\alpha_{\Lambda}$ values). Thus, the dynamics in 1D lattices leads to more extended wave packets and more chaotic  behaviors than in 2D systems. This observation and the analysis of the $T_D / T_L$ ratios, lead to the conjecture that {\it for one and two spatial dimensions there exists a uniform scaling between the wave packet's spreading and its degree of chaoticity}. This can be quantified by assuming
\begin{equation}
 \frac{\Lambda^1 (t)}{m_2^1 (t)} =  \frac{\Lambda (t)}{m_2 (t)},
\label{eq:analytic_mle}
\end{equation}
where the subscript $(^{1})$ refers to 1D systems. To validate this assumption we use Eq.~\eqref{eq:analytic_mle} to estimate the time evolution of $\Lambda (t)$, for both spreading regimes, based on previously obtained numerical results for the MLE of the 1D DKG and DDNLS models~\cite{SGF13,SMS18}, along with the theoretical predictions of~\cite{F10} for the evolution of $m_2$. In particular, Eq.~\eqref{eq:analytic_mle} gives
\begin{equation}
\Lambda \propto  t^{a_m - a_m^1 + \alpha_{\Lambda}^1 },
\label{eq:scaling_of_le_gen}
\end{equation}
resulting to $\Lambda \propto t^{-0.38}$ [$ t^{-0.47}$] for the weak [strong] chaos regime where $a_m=1/5$, $a_m^1=1/3$, $\alpha_{\Lambda}^1=-0.25$ [$a_m=1/3$, $a_m^1=1/2$, $\alpha_{\Lambda}^1=-0.3$], being in very good agreement to $\Lambda  \propto t^{-0.37}$ [$ t^{-0.46}$] observed in Figs.~\ref{fig:sec_mom_mle_weak}(b) and (d) [Figs.~\ref{fig:sec_mon_mle_strong}(b) and (d)].

\begin{figure}
\includegraphics[width=\columnwidth]{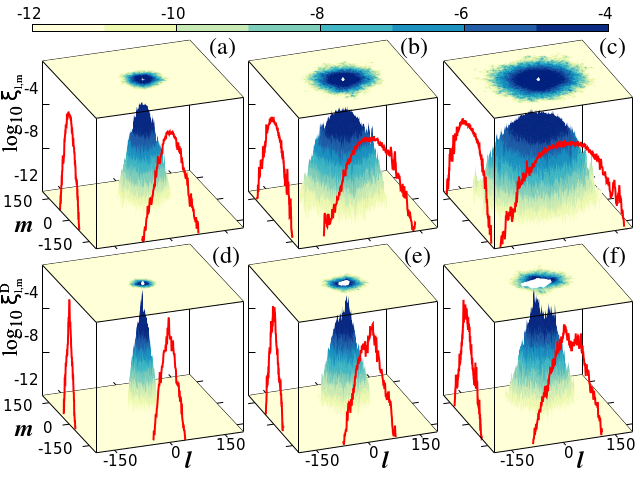}
\caption{
3D density profile and associated 2D color map (upper sides) snapshots of the wave packet $\xi _{l, m}$ [(a)--(c)] and DVD $\xi^D_{l, m}$ [(d)--(f)] for a representative example of the $S2_K$ case at times $\log_{10} t \approx 3.7$ [(a) and (d)], $5.9$ [(b) and (e)] and $7.4$ [(c) and (f)]. The white region at the maps' central part represents the area covered by the distribution's center. The red curves on the sides are distribution's projections along the $l$ and $m$ axes. The color bar at the top is in logarithmic scale.}
\label{fig:energy_norm_dev_vec_distr}
\end{figure}

\begin{figure}
\includegraphics[scale=0.6]{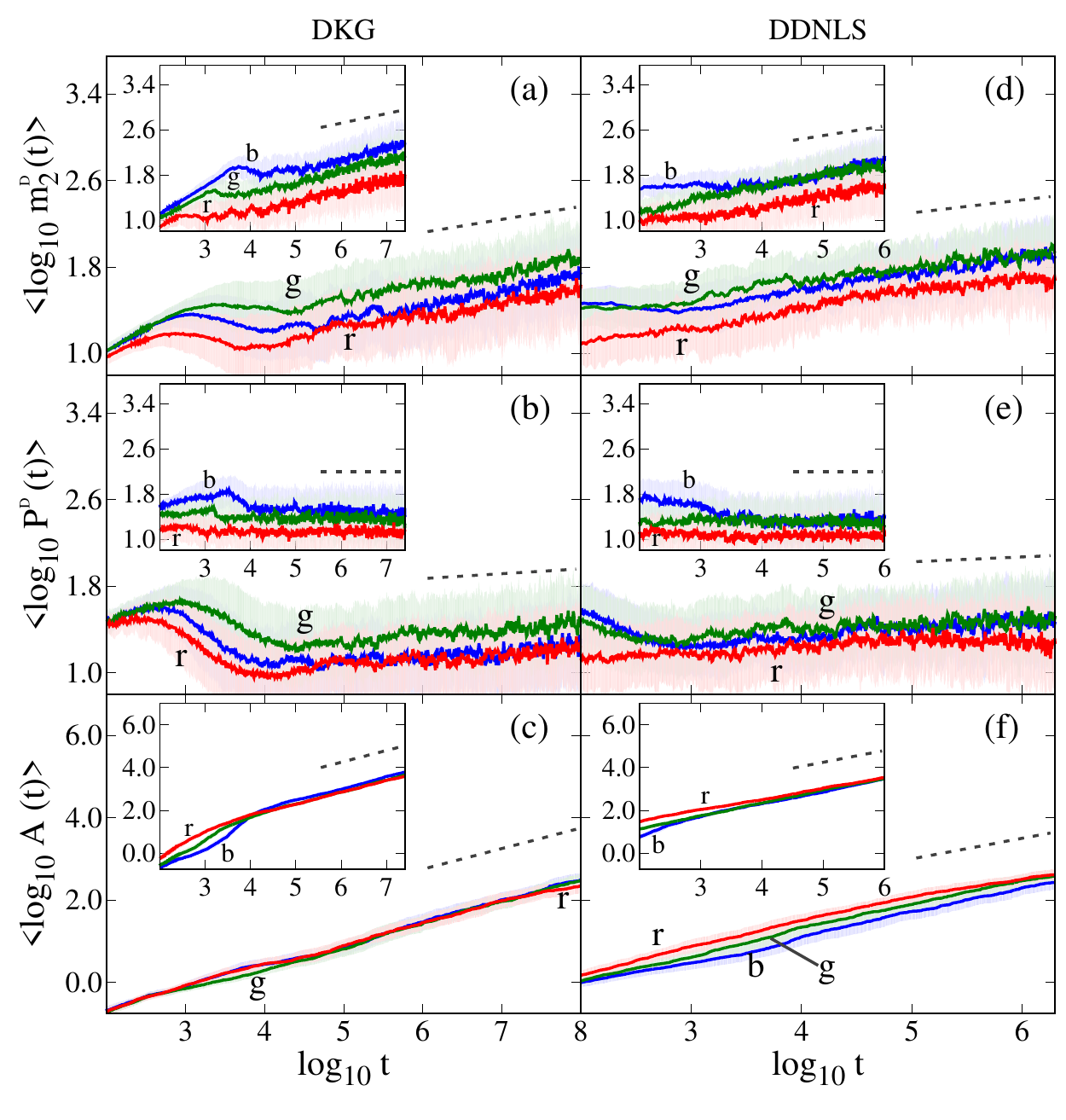}
\caption{\label{fig:DVD_sec_mom_P_array_weak}
DVD characteristics: weak chaos (insets: strong chaos). Averaged results of the  evolution of $m_2^D (t)$ [(a) and (d)], $P^D (t)$ [(b) and (e)], and $A (t)$ [(c) and (f)] for the DKG [(a)--(c)] and the DDNLS [(d)--(f)] model. The curve colors correspond to the cases of Fig.~\ref{fig:sec_mom_mle_weak} (insets: Fig.~\ref{fig:sec_mon_mle_strong}). Shaded areas indicate one standard deviation. The dashed lines denote power law increases with exponents $a_m^D = 0.12$ (insets: $0.17$) [(a) and (d)], $a_P^D = 0.045$ (insets: $0.0$) [(b) and (e)] and $\alpha_A = 0.5$ (insets: $0.55$) [(c) and (f)].}
\end{figure}

The evolution of the DVD associated with the deviation vector $\boldsymbol{w} (t)$ used for the computation of $\Lambda$ has already been implemented to visualize the chaotic behavior of 1D nonlinear lattices and to identify the motion of chaotic seeds, i.e.~regions which are more sensitive to perturbations~\cite{SGF13,SMS18,DVDs_papers}. Local chaotic seeds in 1D disordered nonlinear lattices were also observed and discussed in \cite{OPH09,B11,B12,M14} but, to the best of our knowledge, this is the first time that their behavior is studied in  disordered nonlinear systems with two spatial dimensions. A representative case is shown in Fig.~\ref{fig:energy_norm_dev_vec_distr} where we plot the spatiotemporal evolution of the wave packet $\xi_{l, m}$  [Figs.~\ref{fig:energy_norm_dev_vec_distr}(a)--(c)] and the DVD  [Figs.~\ref{fig:energy_norm_dev_vec_distr}(d)--(f)] for an individual set-up of the $S2_K$ case. We see that the energy density spreads rather symmetrically around the position of the initial excitation, with the distribution's center covering a tiny region around the middle of the lattice [white area at the center of the 2D color maps at the upper sides of Figs.~\ref{fig:energy_norm_dev_vec_distr}(a)--(c)]. On the other hand, the DVD [Figs.~\ref{fig:energy_norm_dev_vec_distr}(d)--(f)] remains always well inside the lattice's excited part, retaining a rather localized character and a concentrated, pointy shape, although its extent increases slightly in time. These behaviors lead, for both  models, to the rather slow increase of the DVD's second moment in Figs.~\ref{fig:DVD_sec_mom_P_array_weak}(a) and (d) as $m_2^D \propto t^{a_m^D}$ with $a_m^D \approx 0.12$ ($0.17$), and the practical constancy of the DVD's participation number $P^D$ in Figs.~\ref{fig:DVD_sec_mom_P_array_weak}(b) and (e) for the weak (strong) chaos regime, with  a very slow increase $P^D \propto t^{0.045}$ observed in  Figs.~\ref{fig:DVD_sec_mom_P_array_weak}(b) and (e).

The chaotic seeds exhibit random fluctuations of increasing width, as the growth of the white region indicating the path traveled by the DVD's center shows in the 2D color maps at the upper sides of Figs.~\ref{fig:energy_norm_dev_vec_distr}(d)--(f). As in 1D lattices~\cite{SGF13,SMS18}, these fluctuations are essential in homogenizing chaos inside the wave packet, supporting in this way its thermalization and spreading. To quantify the area of the region visited by the DVD's center we plot in Figs.~\ref{fig:DVD_sec_mom_P_array_weak}(c) and (f) the evolution of
\begin{equation}
\label{eq:area}
A (t) = R_x (t) \cdot R_y (t),
\end{equation}
where $R_x(t) = \max_{[0, t]} \lbrace \overline{l}^{D} (t)\rbrace - \min_{[0, t]} \lbrace\overline{l}^{D} (t)\rbrace$ and $R_y (t)= \max_{[0, t]}\lbrace\overline{m}^{D} (t)\rbrace - \min_{[0, t]} \lbrace\overline{m}^{D} (t)\rbrace$, in analogy to a similar quantity used in 1D studies (Eq.~(12) of~\cite{SMS18}). In all cases $A \propto t^{\alpha_A}$, with $\alpha_A \approx 0.5$ ($0.55$) for the weak (strong) chaos regime. The larger $A(t)$ and $\alpha_A$ values obtained in the strong chaos case [insets of Figs.~\ref{fig:DVD_sec_mom_P_array_weak}(c) and (f)] clearly indicate the wider and faster motion of chaotic seeds in this regime, where also faster wave packet spreading is observed.

It is worth discussing a bit more the time evolution of $A$ \eqref{eq:area} in connection to the time evolution of $m_2$ \eqref{eq:m2_new}. The wave packet's second moment $m_2$ is only one measure of the wave packet's extent. From its definition in Eq.~\eqref{eq:m2_new} we see that it is measured in units of (distance)$^2$, i.e.~it quantifies the `area' covered by the wave packet. It is important to note that this is actually a weighted measure of that area, with the weight being the energy/norm density $\xi_{l,m}$. In our set-up regions further away from the point of initial excitation are contributing to the $m_2$ value much less as $\xi_{l,m}$ decreases rapidly. On the other hand, the estimator $A$ \eqref{eq:area} of the area visited by the DVD, which is also measured in (distance)$^2$ units, is not weighted and consequently regions further away from the region of the initial excitation are equally contributing to the value of $A$. This is  why the exponents $\alpha_A$ in $A \propto t^{\alpha_A}$, are larger than the exponents $a_m$ in $m_2 \propto t^{a_m}$, something which could create the wrong impression that the area covered by the DVD (quantified by the unweighted quantity $A$) increases faster than the area of the wave packet (quantified by the weighted quantity $m_2$). For example, by comparing Figs.~\ref{fig:energy_norm_dev_vec_distr}(c) and (f) we see that the white region in the 2D color map of Fig.~\ref{fig:energy_norm_dev_vec_distr}(f) (DVD), used for the computation of $A$, corresponds to a region in Fig.~\ref{fig:energy_norm_dev_vec_distr}(c) (wave packet) having $\xi_{l,m}$ values 2-3 orders of magnitude smaller at its boarders with respect to its central part. Thus, in the computation of $m_2$ the outer parts contribute much less, while for $A$ all parts contribute equally.

\section{\label{par:discussion_conclusion} Conclusions}

We conducted a detailed study of the evolution of initially localized wave packets, in both the weak and strong chaos dynamical regimes, of 2D disordered nonlinear lattices by performing long-time and high-precision numerical simulations in large DKG and DDNLS lattices with quartic nonlinearities, completing in this way some previous, sporadic works on this issue~\cite{GS09,LBF12}. We showed the subdiffusive spreading of wave packets resulting in the destruction of AL, and verified the validity of previously made~\cite{F10} theoretical predictions on the characteristics of these spreadings by finding that $m_2 \propto t^{a_m}$ with $a_m \approx 1/5$ ($1/3$) for the weak (strong) chaos regime.

We also investigated the chaotic properties of these systems through the computation of appropriate observables related to their tangent dynamics. The finite time MLE, $\Lambda$, decays in time as $\Lambda \propto t^{\alpha_{\Lambda}}$, with $\alpha_{\Lambda} \approx -0.37$ ($-0.46$) in the weak (strong) chaos case, denoting the decrease of the systems' chaoticity as wave packets spread. Despite this slowing down of chaos, our results show that the chaotization of the lattice's excited part always takes place faster than the wave packet's spreading, i.e.~wave packets first thermalize due to chaos and then spread. Furthermore, no signs of a crossover to regular dynamics is observed, indicating that chaos persists. Conjecturing the similarity of chaotic processes in 1D and 2D systems, along with the existence of a scaling between the wave packet's spreading and chaoticity, which is independent of the lattice's dimensionality [Eq.~\eqref{eq:analytic_mle}], we accurately predict the numerically obtained $\alpha_{\Lambda}$ values. In the future we plan to probe the generality of this conjecture also for 3D systems.

The DVDs' spatiotemporal evolution revealed the mechanisms of  chaotic spreading: localized chaotic seeds oscillate randomly inside the excited part of the lattice, homogenizing chaos in the interior of the wave packet, and supporting in this way its thermalization and subdiffusing spreading. The amplitude of these oscillations increase in time allowing the chaotic seeds to visit all regions of the expanding wave packet. This process is generic as it also appeared in 1D systems~\cite{SGF13,SMS18}.

The fact that in both the DKG and DDNLS models we observed the same evolution laws, with identical numerical exponents for all studied quantities, underline the universality of our findings.

\begin{acknowledgments}
Ch.S.~and B.M.M.~were supported by the National Research Foundation of South Africa. B.S.~was funded by the Muni University AfDB HEST staff development fund. We thank the High Performance Computing facility of the University of Cape Town
and the Center for High Performance Computing for providing their computational resources.
We also thank the  anonymous referees for their comments, which helped us improve the presentation of our work.
\end{acknowledgments}



\begin{thebibliography}{99}
\bibitem{comment_NM} In the cases studied here the smallest considered $W$ value is $W=9$ and the largest $W=14$, which, respectively, correspond to average NM participation numbers of $\approx 60$ and $\approx 10$ (see also the inset of Fig.~1 in~\cite{LBF12}).
\bibitem{com2} In a similar way as in \cite{com1}, from the results of  Figs.~\ref{fig:sec_mon_mle_strong}(a) and (c) we get
    $a_{m} = 0.3309 \pm 0.0001$ ($S1_K$),
    $0.3289 \pm 0.0002$ ($S2_K$),
    $0.3307 \pm 0.0003$ ($S3_K$),
    $0.3292 \pm 0.0002$ ($S1_D$),
    $0.3318 \pm 0.0001$ ($S2_D$) and
    $0.3362 \pm 0.0002$ ($S2_D$).
\end{thebibliography}
\end{document}